%% file: nlse_cap_sub_18_02_25.tex
\documentclass[pre,10pt,aps,twocolumn]{revtex4-2}

\input{macros_alm}

\newcommand{\SnoProp}{\overline{{\cal S}}_{noProp}}

\newcommand{\SNR}{\mbox{SNR}}
\newcommand{\SE}{\mbox{SE}}

\usepackage{comment}
\usepackage[textwidth=30mm]{todonotes}

\setstcolor{red}
\sethlcolor{SkyBlue}

\definecolor{lightskyblue}{rgb}{0.53, 0.81, 0.98}

\begin{document}


\title{Spectral Efficiency Expression for the Non-Linear Schr\"{o}dinger Channel\\ in the Low Noise Limit Using Scattering Data}

\author{Pavlos~Kazakopoulos$^{1}$  and Aris~L. Moustakas$^{1,2}$}
\affiliation{$^{1}$Department of Physics, National \& Kapodistrian University of Athens, Greece \\
$^{2}$Athena Research Center / Archimedes Research Unit, Athens, Greece}



\begin{abstract}
Transmission through optical fibers offers ultra-fast and long-haul communications. However, the search for its ultimate capacity limits in the presence of distributed amplifier noise is complicated by the competition between wave dispersion and non-linearity. 
In this paper, we exploit the integrability of the Nonlinear Schr\"{o}dinger Equation, which accurately models optical fiber communications, to derive an expression for the spectral efficiency of an optical fiber communications channel, expressed fully in the scattering data domain of the Non-linear Fourier Transform and valid in the limit of low amplifier noise. We utilize the relationship between the derived noise-covariance operator and the Jacobian of the mapping between the signal and the scattering data to obtain the properties of the former. Emerging from the structure of the covariance operator is the significance of the Gordon-Haus effect in moderating and finally reversing the increase of the spectral efficiency with power. This effect is showcased in numerical simulations for Gaussian input in the high-bandwidth regime.
\end{abstract}


\maketitle


\section{Introduction}
\label{sec:Introduction}
In recent years, the looming capacity crunch \cite{Tkach2010_ScalingCrunch, 8610949} has renewed the interest in increasing optical fiber throughput. 
In contrast to linear communications channels with additive noise, whose capacity limits were established in the 1940's, the ultimate theoretical limits of optical fiber throughput are still unknown. This is of course due to the non-linearities inherent in optical fiber transmission, which greatly complicate the dynamics of the signal.  In the propagation of light through silica fiber, increasing transmission power leads to the emergence of the Kerr nonlinearity due to four-wave mixing. This can be modelled by an intensity-dependent index of refraction $n(\omega,A)\approx n_0(\omega)+n_2 |A|^2$, where $A$ is the signal amplitude. The propagation of a single polarization mode through the fiber is adequately described by the nonlinear Schr\"{o}dinger equation (NLSE) \cite{Agrawal1995_book_NonlinearFOptics}
\begin{eqnarray}
i\frac{\partial A}{\partial z} + \frac{\beta_2}{2}\frac{\partial^2 A}{\partial t^2} + \gamma |A|^2 A = n(z,t)
\label{nlse_def}
\end{eqnarray}
For light propagation inside a fiber, $z$ in \eqref{nlse_def} denotes position along the fiber and $t$ denotes time in the co-moving frame. The parameters $\beta_2$ and $\gamma$ are related to the group velocity dispersion (GVD) and the Kerr nonlinearity, respectively \cite{Agrawal1995_book_NonlinearFOptics, Glass_et_al}, while $n(z,t)$ is the additive amplifier noise.
Currently, the most widely used approach, Wavelength Division Multiplexing (WDM), works with minimizing the effects of channel dispersion by spreading the signal over different frequency bands. WDM is optimal in the absence of non-linearity but it breaks down for high transmission powers, where nonlinear effects become too strong to ignore.

\subsection{Summary  of Prior Work} 

We briefly summarize the increased recent research interest in the study of the spectral efficiency in the presence of nonlinearity. In \cite{narimanov2002channel}, the strength of the non-linearity was treated perturbatively. In \cite{Mitra_Stark_2001, kahn2004spectral,chen2010closed} the optical fiber throughput was analyzed using WDM and an upper input power limit was found, beyond which the throughput tends to decrease due to increased interference. A different approach was taken by \cite{turitsyn2003information, kramer2018autocorrelation, keykhosravi2019accuracy, reznichenko2020path, fahs2021capacity}, where the dispersion term in (\ref{nlse_def}) was set to zero and only the nonlinearity was taken into account. In this limit, the spectral efficiency increases for large powers as $\frac{1}{2}\log(\SNR)$, i.e. half the value of the spectral efficiency of the linear complex channel. However, the bandwidth over which zero dispersion is feasible is not large and hence this approximation has limited scope. A number of other methods to estimate the fundamental limits of fiber-optical communications in the presence of nonlinearity have been proposed \cite{ellis2017performance, ip2009fiber, bosco2011analytical, secondini2012analytical, secondini2013achievable}. There have also been direct approaches using the NLSE \cite{essiambre2008capacity, essiambre2010capacity, essiambre2012capacity, mecozzi2012nonlinear}, and these are currently the state of the art.

More recently, a remarkable property of the NLSE has received increased attention in connection with these efforts: Despite its nonlinearity, the NLSE in the absence of the noise term is {\it integrable}, i.e. it is exactly solvable for arbitrary initial conditions by means of a nonlinear transformation known as the Inverse Scattering Transform (IST) \cite{Zakharov_Shabat_1972, AKNS1974, Konotop_book}, or as the Nonlinear Fourier Transform (NFT) in recent optical fiber channel literature \cite{yousefi2014information,  turitsyn2017nonlinear}. The NFT transforms the original nonlinear partial differential equation that describes the dynamics of the input signal into a set of canonical variables, akin to action-angle variables, that have simple linear dynamics. The message can be encoded using these variables, and then decoded at the receiver using the inverse NFT. To this end, fast algorithms have been recently developed that allow NFT-encoding and decoding in near-linear time \cite{wahls2015fast, chimmalgi2019fast, vaibhav2018higher}. In addition, several studies, both analytical and experimental, have demonstrated ways in which the various non-linear modes can be modulated and processed and have also obtained achievable information rates \cite{prilepsky2014nonlinear, hari2016multieigenvalue, zhang2016achievable, le201764, frumin2017new, gui2018nonlinear, leible2020back, aref2018modulation, solitons_2000km, Xianhe2019_b_mobulation, rademacher2021peta, Kodama2023_HyperMultiLevel, Turitsyn2023_BrightProspects}. 

Other methods based on the NFT have also been proposed  taking into account multi-soliton solutions \cite{cartledge2017digital, perturbations_nfdm, le2015nonlinear, yousefi2019linear, Suret2024_SolitonGas, KaurDhawanGupta2023_ReviewLongHaul}. Recently, a number of papers have studied the impact of continuous (non-solitonic) nonlinear modes of the NLSE and have obtained bounds on their spectral efficiency \cite{tavakkolnia2017capacity, derevyanko2021channel, derevyanko2016capacity}. However, a similar analysis has not yet been made taking both continuous and solitonic modes of the NLSE into account.

\subsection{Contributions}
\label{sec:Contributions}

In this paper, we provide an analysis of the spectral efficiency of the NLSE taking advantage of its integrability. After highlighting the importance of permutation errors for the solitonic degrees of freedom when the noise is non-negligible, we derive an expression of the spectral efficiency in the domain of the scattering data when amplifier noise is low. We analyze the covariance operator of the noise and its relation to the Jacobian of the canonical transformation between the signal and its scattering data, which allows us to re-derive the Shannon upper-bound of the spectral efficiency \cite{yousefi2015upper} within this framework. The structure of the covariance operator showcases the importance of the Gordon-Haus effect \cite{Gordon_Haus_1986} in reducing the performance of the system. Finally, by applying our method to the special case of a white Gaussian input signal distribution, which has a known distribution of eigenvalues and scattering data \cite{Kazakopoulos2008_NLSE}, we calculate the spectral efficiency as a function of input SNR.

\subsection{Paper Outline}
\label{sec:Outline}

In Section \ref{sec:InverseScatteringTransform} we introduce the noise model and arrive at a dimensionless form of the noisy NLSE. We introduce the Zakharov-Shabat operator and the scattering data that emerge from it, as well as the structure of their perturbation under additive noise. In Section \ref{sec:CovarianceMatrix} we express the covariance matrix of the noise in terms of the scattering data and discuss its properties, while in Section 
\ref{sec:SolitonIT} we derive the formula for the spectral efficiency and obtain two bounds, namely the Shannon upper bound and a useful lower bound. Section \ref{sec:GaussianInputCase} introduces the Gaussian input case and its properties in the high-bandwidth limit and in Section \ref{sec:Numerics} we describe the numerical methodology and discuss the resulting spectral efficiency in the case of white Gaussian input. Section \ref{sec:Conclusion} concludes  the paper. Appendices \ref{sec:PertTheory} and \ref{sec:ProofMIBounds} provide details for Sections \ref{sec:CovarianceMatrix} and \ref{sec:SolitonIT} respectively.

\section{Model Description and Inverse Scattering Transform}
\label{sec:InverseScatteringTransform}

\subsection{Signal and Optical Fiber Channel Model}
\label{sec:ChannelModel}

We first define the model for the amplification noise affecting the fiber communications system. Specifically, we consider additive noise injected at $K$ equally spaced amplifiers that offset the dissipation of the electric field amplitude propagating along the fiber. These add noise to the signal, mainly because of amplified spontaneous emission of photons (ASE) \cite{Agrawal1992_book_FOCommunicationSystems, Kaminow_Koch_book}. If the distance between successive amplifiers is $L$ then the noise variance may be estimated as $\sigma^2=n_{sp}E_{ph}(g-1)$ \cite{Essiambre2010_CapacityLimitsOpticalFiberNetworks}, where $E_{ph}$ is the photon energy, $n_{sp}$ is the emission factor, and $\log_{10}g=0.1\alpha_{loss} L$ where $\alpha_{loss}$ is the loss factor. Typical values for these parameters can be found in Table \ref{tab:parameters}.
To model these noise effects, the NLSE must  modified by an additive random term $n(z,t)$ of the form:
\begin{equation}\label{noise_def}
n(z,t) = i\sigma\sum_{k=1}^K w_k(t) \delta(z-kL)
\end{equation}
where $w_k(\cdot)$ is the noise inserted at the $k$th amplifier, assumed to be bandlimited with bandwidth $B$. 
We rescale the variables in \eqref{nlse_def} to make the model dimensionless as follows: Define $A=\sqrt{R} u$, $z=\ell x$ and let $t\to t_s t$. If we relate them by $\ell^{-1}=\frac{\beta_2}{2t_s^2}=\frac{\gamma R}{2}$, in rescaled units the noise variance becomes $\epsilon^2=\sigma^2/(Rt_s)$, the distance between amplifiers is $L_s=L/\ell$, while the signal duration is $T_s=T/t_s$. We still have one free variable, $R$, which we shall set at a later stage. In the new variables, the equation of propagation is:
\begin{eqnarray}
i\frac{\partial u}{\partial x} + \frac{\partial^2 u}{\partial t^2} + 2 |u|^2 u = i\epsilon\sum_{k=1}^K w_k(t) \delta\left(x-kL_s\right)
\label{nlse_rescaled}
\end{eqnarray}
The noise is added to the signal at distances $kL_s$, adding successive noise distortions of the form $\delta u(kL_s,t)=w_k(t)$. We also assume that the input pulse has temporal duration $T$, and the same bandwidth $B$ as the noise (i.e. the frequency spectrum is bounded by $|f|\leq B/2$).  Hence, in rescaled units, 
\begin{equation}\label{w-corrs}
  \ex\left[w_k(t)w_{k'}^*(t')\right]= \epsilon^2\delta_{kk'}\delta_{Bt_s}(t-t')
\end{equation}
where $\delta_Q(t)=Q\, \text{sinc}(Qt)$ \cite{yousefi2015upper} ($\text{sinc}(x)=\sin(x)/x$). The function $\delta_Q(t)$, which tends to the Dirac $\delta(t)$ when $Q\to\infty$, corresponds to a hard cutoff for the frequency bandwidth -- a different filter will provide a different expression for $\delta_Q(t)$. 

To make analytical progress, we shall focus on the low-noise limit $\epsilon\ll 1$ and treat $\epsilon$ as a perturbation expansion parameter.
Given the signal bandwidth $B$ and the duration $T$, Nyquist's theorem states that the maximum number $M$ of independent complex degrees of freedom of the signal is $M=BT$. 
In order to be able to neglect the effects at the boundary and the interference with adjacent signals, we shall assume a long signal duration $T$ and take the limit $M\to \infty$.
Since both signal and noise have the same bandwidth, it makes sense to discard higher frequencies from the analysis, as discussed in \cite{Kramer_bound}, or, equivalently, discretize time at the inverse (normalized) bandwidth $\tau=(Bt_s)^{-1}$. 

Furthermore, we impose an average signal power constraint, i.e.
\begin{align}\label{power_constraint}
\ex\left[|A(t)|^2\right] = P
\end{align}
and assume incoming signals that have the maximum degree of statistical independence, in order to maximize the input entropy. 
Therefore, we can write 
\begin{equation}\label{u-corrs}
  \ex\left[u(t)u^*(t')\right]= D\delta_{Bt_s}(t-t'),
\end{equation}
where $D=\frac{P}{RBt_s}$ is the variance of the $M=BT$ complex independent degrees of freedom in the frequency domain. Note that at this point the above covariance constraint does not necessarily imply Gaussianity of the signal. However, we will now fix the values of $R,t_s$ such that $D=1$. and assume that $Bt_s\gg 1$, so that we may assume that both the signal and the noise are $\delta$-correlated. This is the same regime that was analyzed in \cite{Kazakopoulos2008_NLSE}, providing a nontrivial distribution of solitons and allows us to take the continuum-time limit.


\subsection{Scattering Data of NLSE}
\label{sec:scattering_data_intro}

The integrability of the NLSE in the absence of noise means that the initial pulse $u(x=0,t)\equiv u(t)$ can be expressed in terms of the scattering data of the associated linear Zakharov-Shabat operator:
\begin{eqnarray}
\bU \bPsi(t)= \left( \begin{array}{lr} i\frac{\partial}{\partial t} & u^*(t) \\ - u(t) & -i\frac{\partial}{\partial t} \end{array} \right) \bPsi(t) = \lambda\bPsi(t)
\label{zs}
\end{eqnarray}
where $\bPsi(t)=[\psi_1(t),\psi_2(t)]^T$ are two-component complex eigenfunctions. The operator $\bU$ is not Hermitian and so its eigenvalues $\lambda=\xi+i\eta$ are in general complex. As we shall see below, the evolution of the scattering data in terms of the distance of propagation $x$ in the absence of noise is trivial \cite{AKNS1974, Konotop_book}. 

In the complex spectrum sector, the scattering data consist of the discrete complex eigenvalues $\lambda_n$, with $n\in \N$, which remain constant under propagation, and the $b_n$, defined as
\begin{eqnarray}
\log b_{n}=\log \lim_{t\to\infty}\frac{\psi_{2n}(t)}{\psi_{1n}(-t)},
\label{b_z_def}
\end{eqnarray}
which are related to the ``center'' of the localized eigenfunction $\bPsi_n$ and have a simple dependence on propagation distance $x$:
\begin{eqnarray}\label{logbn_deterministic_variation}
\log b_n(x)=\log b_n(0)-4i\lambda_n^2 x.
\end{eqnarray}
The eigenvalues $\xi$ located on the real axis ($\eta=0$) on the other hand form, in the infinite pulse duration limit, a continuum of extended eigenstates. The corresponding scattering data is the complex reflection coefficient $\rho_\xi$, whose logarithm also depends linearly on $x$:
\begin{eqnarray}\label{logrho_deterministic_variation}
\log \rho_\xi(x)=\log \rho_\xi(0)-4i\xi^2 x.
\end{eqnarray}
The ZS operator has two discrete symmetries, which are important when counting independent degrees of freedom of the pulse $u(t)$. First, there is an involutive symmetry as $\bar{\bPsi}_n=[\psi_{2n}^*,-\psi_{1n}^*]^T$ is the eigenfunction for $\lambda_n^*$. A second symmetry becomes apparent when we discretize the differential equation, for example following the modified Ablowitz-Ladik recipe \cite{Weideman_Herbst_Modified_Ablowitz_Ladik}, so that time takes discrete values $t=p\tau$, where $p\in \N$ and $\tau$ is the discrete time-step. In this setting, the real part of the eigenvalue $\xi$ takes values in the interval $\xi\in \left(-\frac{\pi}{\tau}, \frac{\pi}{\tau}\right)$. However, for any two eigenvalues $\lambda_n$, $\lambda_m$ with equal imaginary parts and real parts separated by $|\lambda_n-\lambda_m|=\frac{\pi}{\tau}$, the discrete eigenfunctions are related by: 
\begin{equation}\label{discreteSymmetry_Psi}
\left[
\begin{array}{c} 
\psi_{1m}(p\tau) \\ \psi_{2m}(p\tau) 
\end{array} 
\right] \leftrightarrow (-1)^p\left[ 
\begin{array}{c} 
\psi_{1n}(p\tau) \\ -\psi_{2n}(p\tau) 
\end{array} 
\right] 
\end{equation}
This symmetry is a result of from the fact that in the discrete NLSE the eigenvalues $z_n=e^{-i\lambda_n\tau}$ come in pairs $\pm z_n$ (see Section 3.2.2 in \cite{Ablowitz2004_DNLS_book}). 
In conclusion, the eigenvalues with $\eta\geq 0$ and $\xi>0$ together with their corresponding scattering coefficients will carry all the information contained in the initial pulse $u(t)$.

A simple counting argument for why the scattering data $b_n$, $\lambda_n$ and $\rho_\xi$ are sufficient to fully describe the variations of the incoming pulse can be made if we count the degrees of freedom of the system in a discretized setting. 
Specifically, if we express the incoming complex signal using $M$ discrete points in time, the discrete ZS operator becomes a $2M\times 2M$ matrix, hence having $2M$ eigenvalues. As we saw, the complex eigenvalues come in quadruplets, i.e. $\lambda$, $\lambda^*$, $\lambda+\pi/\tau$, $\lambda^*+\pi/\tau$, following the symmetries of the ZS matrix described above. Denoting as $N$ the number of the eigenvalues in the upper right quadrant of the $\lambda$ complex plane, we see that each has 2 complex (4 real) independent degrees of freedom, corresponding to $\log b_n$ and $\lambda_n$. The (possibly) remaining $2N_{c}=2M-4N$ real eigenvalues come in pairs, $\xi$ and $\xi+\pi/\tau$, with one complex (and hence two real) scattering coefficient $\rho_\xi$ for each pair. It is reasonable to assume that the $M$ complex scattering data values are approximately independent, since they correspond to orthogonal eigenstates. Thus, the set of scattering data contains the same number of independent degrees of freedom as in the original discretized picture of the incoming signal in the time domain. Since from Nyquist's theorem we know that the incoming signal has $BT$ complex ($2BT$ real) degrees of freedom, we expect that a discretization of $M\geq BT$ points with normalized time step $\tau=(Bt_s)^{-1}\ll 1$ will be sufficient to describe the information content of the signal. In addition, in the limit that $\tau$ is very small, we can take the continuum time-limit of the equations describing the dynamics. This implicitly assumes that all degrees of freedom at smaller time-scales do not contribute in the information transfer.

\subsection{Impact of Noise on Scattering Data}
\label{sec:ImpactNoiseScatData}

To find how noise affects the scattering data, we express the  variations in the scattering data due to an infinitesimal (and local in space) variation $\delta u(t)$ in the initial pulse as \cite{Kaup2010_SERevisited}:
\begin{align}\label{dlambda_du}
\delta\lambda_{n}^{0} = \int dt \frac{ 
\psi_{2n}^{2}(t)\delta u^{*}(t) -  \psi_{1n}^{2}(t) \delta u(t)}{\gamma_n^{}}
\end{align}
with $\gamma_n$ being the normalization constant corresponding to the eigenstate, defined as $\gamma_n=2\int dt \psi_{1n}(t)\psi_{2n}(t)$. The variation of $\rho_\xi$ is given by
\begin{align}\label{drho_du}
\delta\rho_\xi^{0} = \frac{\int dt 
\left( \phi_{1\xi}^2(t)\delta u(t) -  \phi_{2\xi}^2(t)\delta u^{*}(t)\right)}{-ia(\xi)^2}
\end{align}
where $a(\xi)$ is a scattering data coefficient \cite{AKNS1974, Kaup2010_SERevisited} reflection coefficient evaluated at $\lambda=\xi$. Finally,  the local variation of $b_n$ can be expressed in terms of the canonical variable $\mu_n=\log\left(\frac{b_n}{a'_n}\right)$, where $a'_n=a'(\lambda_n)$, as
\begin{align}
\delta\mu_n^{} &= \int dt \frac{\left(\psi_{2n}^2(t)\right)^{'}\delta u^{*}(t) 
-  \left(\psi_{1n}^2(t)\right)^{'} \delta u(t)}{\gamma_n^{}} -\frac{a''_n}{a'_n} \delta\lambda_n^0
\nonumber \\
&\equiv \delta\mu^0_n -\frac{a''_n}{a'_n}\delta\lambda_n^0
\label{dlogb_du}
\end{align}
where $a''_n=a''(\lambda_n)$ and $\left(\psi_{in}^2(t)\right)^{'}=2\psi_{in}(t)\psi^{'}_{in}(t)$ for $i=1,2$ and $\psi^{'}_{in}(t)$ are the solutions of the derivative of \eqref{zs} with respect to $\lambda$, evaluated at $\lambda=\lambda_n$ and $\bPsi=\bPsi_n$. As we shall see, this last term will have no impact in any of the spectral efficiency calculations because it is linearly dependent on the variations of the eigenvalues and can be evaluated at the initial location to leading order.

The coefficients of $\delta u(t)$ and $\delta u^*(t)$ in the expressions above comprise the Jacobian of the transformation from $\bLambda = \left(\{\lambda_n\}, \{\mu_n\}\right)$ and $\brho=\{\rho_\xi\}$ to $\{u(t), u(t)^*\}$. Denoted for compactness by $J_{\lambda u}$, $J_{bu}$, $J_{\rho u}$, $J_{\lambda \bar{u}}$ etc., it corresponds to variations with respect to $u(t)$ and $u^*(t)$ respectively, e.g. $\left[J_{\lambda u}\right]_{n,t}=\delta\lambda_n/\delta u(t)$. Due to the integrability of the NLSE, the above infinitesimal transformation is canonical in the Hamiltonian sense \cite{Faddeev2007_HamiltonianMethods}, so its Jacobian has unit norm determinant. The inverse of this Jacobian operator, $\bar{J}$, can also be expressed in a similar way using the following expansion \cite{Kaup2010_SERevisited}:
\begin{align}\label{Jacobian_u_lambda}
\delta u = &-\int_0^\infty \frac{d\xi}{i\pi}\left(\phi_{2\xi}^2\delta\rho_\xi^{}+ \phi_{1\xi}^{*2}\delta\rho_\xi^*\right) 
\\ \nonumber
&+2\pi i \sum_n\left(\frac{\left(\psi_{2n}^{2}\right)'}{\gamma_n^{}} \delta \lambda_n^{} 
+ \frac{\left(\psi_{1n}^{*2}\right)'}{\gamma_n^*} \delta \lambda_n^* \right)
\\ \nonumber
&+2\pi i \sum_n\left(\frac{\psi_{2n}^2}{\gamma_n^{}} \delta \mu_n^{} 
+ \frac{\psi_{1n}^{*2}}{\gamma_n^*} \delta \mu_n^* \right)
\end{align}

\section{Covariance Matrix of Additive Gaussian Noise}
\label{sec:CovarianceMatrix}

\subsection{Introduction of Noise}
\label{sec:Noise_Intro}

The above analysis allows us to obtain the leading correction to the scattering data due to the additive amplifier noise.
In the presence of noise, \eqref{nlse_rescaled} is no longer integrable and the scattering data of the equation (namely $\{\lambda_n\}$, $\{\mu_n\}$ and $\{\rho_\xi\}$) become spatially varying. Since the noise is injected into the signal at discrete spatial intervals, as seen in \eqref{noise_def}, the total variation of the scattering coefficients can be expressed as a sum of such individual variations, each in the form of  \eqref{dlambda_du}, \eqref{drho_du}, \eqref{dlogb_du}. Hence, for the eigenvalues $\lambda_n$ and reflection coefficients $\rho_\xi$ we have
\begin{align}\label{sum_delta_lambda_nk}
\delta\lambda_{n}^0 = \sum_{k=1}^K \delta\lambda^{0}_{n,k} \\ 
\delta\rho_\xi^0 = \sum_{k=1}^K \delta\rho^{0}_{k\xi}
\end{align}
where each $\delta\lambda^{0}_{n,k}$ and $\delta\rho^{0}_{\xi,k}$ are of the form \eqref{dlambda_du}, \eqref{drho_du}, respectively, with $\delta u=\epsilon w_k(t)$, and $\phi_{i\xi,k}(t)=\phi_{i\xi,k}(t,x=kL_s)$, $\psi_{in,k}(t)=\psi_{in}(t,x=kL_s)$, where $i=1,2$,  evaluated, to leading order in the absence of noise, at location $x=kL_s$ (for $k=1,\ldots,K$) in the fiber, i.e. with input signal $u(t,kL_s)$.

The variation of each $\log b_n$ (and hence its corresponding $\mu_n$) is complicated by the fact that it includes a deterministic variation component which depends on its eigenvalues, as seen in \eqref{logbn_deterministic_variation}. In this case, the variation of the latter at each amplifier  will result to additional fluctuations of the former. Specifically,
\begin{align}\label{delta log bn}
\delta \mu_{n} &= \sum_{k=1}^K \left(\delta \mu^0_{n,k}-\frac{a''(\lambda_{n,k})}{a'(\lambda_{n,k})}\delta \lambda^0_{n,k}\right) \\ \nonumber
&- 4i\sum_{k=1}^K \left(\lambda_{n,k-1}^2-\lambda_{n,0}^2\right)L_s 
\end{align}
 The additional term in the second line represents the so-called Gordon-Haus effect \cite{Gordon_Haus_1986}, resulting from the variations in the soliton velocities. It is important to point out that no such term appears in the continuous (real) spectrum sector \cite{derevyanko2016capacity}. This is because the real eigenvalues correspond to delocalized eigenfunctions, which are not sensitive to random perturbations. In other words, the equation corresponding to \eqref{dlambda_du} for real eigenvalues gives a vanishing contribution, due to the extensivity of the eigenfunctions.

It is convenient to define $N$-dimensional vectors $\delta \blambda^0_k$, $\delta \bmu_k^0$, and the $N_c=M-2N$-dimensional vector $\delta \brho^0_k$ (together with their corresponding complex-conjugates), for the corresponding variations $\delta \lambda^0_{n,k}$, $\delta \mu^0_{n,k}$ etc. In this notation, to leading order in $\epsilon$, \eqref{delta log bn} takes the compact form 
\begin{align}
\label{deltalogbn3}
\delta \bmu_k = \delta\bmu_k^0 - \left(i\alpha_k \bDelta_\lambda+\bA_0\right)\delta \blambda_k^0
\end{align}
where $\alpha_k=8(K-k)L_s$ and $\bDelta_\lambda$, $\bA_0$ are $N\times N$ diagonal matrices with $\lambda_{n,0}$, $\frac{a''(\lambda_{n,0})}{a'(\lambda_{n,0})}$ as their $n$th element, respectively. 

Since the noise between amplifiers is independent, the full noise covariance matrix ${\cal S}$ can be expressed, to leading order in the noise, as follows:
\begin{equation}\label{S_mat_def}
{\cal S}=\sum_{k=1}^K {\cal S}_k =\epsilon^2\sum_{k=1}^K {\cal M}^k
\end{equation}
where ${\cal M}^k$ is the normalized covariance matrix due to the noise injected at the $k$th amplifier.

\subsection{Properties of ${\cal S}$}
\label{sec:PropertiesOfS}

In the remainder of this section we shall discuss a number of important properties of ${\cal S}$, which be crucial in determining the spectral efficiency. Based on the above analysis, ${\cal M}^k$ has two parts. The first, denoted by ${\cal M}^{0k}$, corresponds to the noise injected in the absence of the term proportional to $\delta \blambda$ in \eqref{deltalogbn3} and can be expressed as 
\begin{align}
{\cal M}^{0k} &= 
\left[
  \begin{array}{ccc}
    {\cal M}_{\lambda\lambda}^{0k} & {\cal M}_{\lambda \mu}^{0k\dagger} & 
    {\cal M}_{\lambda\rho}^{0k\dagger} \\
    {\cal M}_{\lambda \mu}^{0k} & {\cal M}_{\mu \mu}^{0k}  &
    {\cal M}_{\rho\mu}^{0k\dagger} \\
    {\cal M}_{\lambda \rho}^{0k} & {\cal M}_{\rho\mu }^{0k}  &
    {\cal M}_{\rho \rho}^{0k}
  \end{array}
\right]
\end{align}
where each block denotes the covariance between the two types of variations appearing as subscripts. For example, ${\cal M}_{\lambda\rho}^{0k}$ is the covariance matrix between $\delta\blambda$ and $\delta\brho$, etc. Explicit expressions of each block in terms of the eigenfunctions, as seen in \eqref{dlambda_du}, \eqref{drho_du}, \eqref{dlogb_du}, appear in Appendix \ref{sec:PertTheory}. From the above analysis, it can be seen that ${\cal M}^{0k}$ can be written compactly in terms of the Jacobian $J_k$ of the transformation  $\left(\{\lambda_n\},    \{\mu_n\},\{\rho\}\right)\rightarrow (u, u^*)$ evaluated at amplifier $k$, as follows 
\begin{equation} \label{eq:M=JJ}
{\cal M}^{0k} = J_kJ_k^\dagger
\end{equation}
Hence, if we neglect the second term in \eqref{deltalogbn3} and assume that there is only a single amplifier present, the determinant of the noise, to leading order, is equal to 
\begin{align}
    \det({\cal S}_k)=\epsilon^{2M}\det({\cal M}^k)=\epsilon^{2M}
\end{align}
where we have used \eqref{eq:M=JJ} and the fact that the determinant of the Jacobian matrix has unit modulus. Hence, the determinant of the covariance of the noise of a single amplifier in the absence of the second term in \eqref{deltalogbn3} is equal to that of the noise of a linear system with additive Gaussian noise and is transparent to the nonlinearity. This is a direct consequence of the fact that the transformation between the addition of infinitesimal noise and the corresponding variations in the  scattering data is canonical, and thus volume preserving. 

The second part of the covariance matrix ${\cal M}^k$ is the contribution of the second term in \eqref{deltalogbn3}, denoted ${\cal G}^k$ and defined by:
\begin{equation} \label{eq:Mk_def}
{\cal M}^{k} = {\cal M}^{0k} +{\cal G}^k
\end{equation}
so that ${\cal M}^k$ takes the following suggestive form
\begin{align}
{\cal M}^{k} &= 
\left[
  \begin{array}{cc}
    {\cal M}_{\lambda\lambda}^{0k} &  {\cal Z}^{k\dagger}
    \\
    {\cal Z}^{k}
    &
    {\cal X}^k
  \end{array}
\right]
\end{align}
where 
\begin{align}\label{eq:Zk_def}
{\cal Z}^{k} &= 
\left[
    \begin{array}{cc}
    {\cal M}_{\lambda \mu}^{0k}-{\cal D}^k{\cal M}_{\lambda\lambda}^{0k},  &
    {\cal M}_{\lambda \rho}^{0k} 
\end{array}
\right]^T,
\end{align}
\begin{align}\label{eq:Xk_def}
{\cal X}^k &= 
\left[
  \begin{array}{cc}
    {\cal M}_{\mu \mu}^{0k}  &  {\cal M}_{\rho\mu}^{0k\dagger} \\
    {\cal M}_{\rho\mu }^{0k} &   {\cal M}_{\rho \rho}^{0k}
  \end{array}
\right] +
{\cal Z}^k
\left[{\cal M}_{\lambda\lambda}^{0k}\right]^{-1}
{\cal Z}^{k\dagger}
\\ \nonumber
&-
\left[
\begin{array}{c}
{\cal M}_{\lambda \mu}^{0k} \\
{\cal M}_{\lambda \rho}^{0k} 
\end{array}
\right]
\left[{\cal M}_{\lambda\lambda}^{0k}\right]^{-1}
\left[
\begin{array}{c}
{\cal M}_{\lambda \mu}^{0k} \\
{\cal M}_{\lambda \rho}^{0k} 
\end{array}
\right]^T,
\end{align}
and 
\begin{align}\label{eq:Dk_def}
{\cal D}^k &=  \left[
  \begin{array}{cc}
    i\alpha_k \bDelta_\lambda+\bA_0 & {\mathbf 0} \\
    {\mathbf 0}  &
    -i\alpha_k \bDelta_\lambda^*+\bA_0^*
  \end{array}
\right]
\\ \nonumber
&\triangleq\alpha_k {\cal D}_1 + {\cal D}_0
\end{align}
The first andarguably more important term, denoted ${\cal D}_1$ and resulting from the last term in \eqref{delta log bn}, encodes the Gordon-Haus effect, which produces random shifts in the velocities of the solitons. The second term, ${\cal D}_0$, derives from the second term in \eqref{delta log bn} and as we shall see, it will cancel completely from the final result. 
For general matrices $\bA,\bB,\bC,\bD$, using the identity $\det \left[
  \begin{array}{cc}
    \bA &\bB \\
    \bC  & \bD
    \end{array}
\right]=\det(\bA)\det(\bD-\bC\bA^{-1}\bB)$
we find that 
\begin{align}
    \det\left({\cal M}^k\right)&= \det\left({\cal M}_{\lambda\lambda}^{0k}\right)
    \det\left({\cal X}^k-{\cal Z}^k
\left[{\cal M}_{\lambda\lambda}^{0k}\right]^{-1}
{\cal Z}^{k\dagger}\right)
\nonumber \\
&=\det\left({\cal M}^{0k}\right)=1
\end{align}
In the last line, we have used \eqref{eq:M=JJ} and the fact that  the Jacobian $J_k$ has unit norm determinant. Hence, for the case of a single amplifier, the second term of \eqref{deltalogbn3}, and correspondingly ${\cal G}^k$ in \eqref{eq:Mk_def}, does not play a role. The latter simplification corresponds to the fact that no matter where the amplifier is located, the receiver can backpropagate the signal to the signal just out of the amplifier, where the Gordon-Haus effect is negligible. 

When there are more than one amplifiers in the system, this backpropagation is no longer possible. Since the Gordon-Haus terms are distance-dependent (corresponding to injections at different locations, with different $\alpha_k$), the above simplification cannot be applied. In addition, there is a summation over ${\cal M}^{0k}$, which cannot be reduced to a single product of Jacobian matrices. Nonetheless, because the term $\bA_0$ is independent of $k$, using the same matrix identity mentioned above, it can be shown to cancel out from the determinant of the full-covariance matrix.

Concluding this section, we see that while in the presence of a single amplifier the noise has the same form as additive Gaussian noise in a linear channel, the presence of multiple non-colocated amplifiers complicates the form of the noise in two ways. The first is the appearance of a sum of ${\cal M}^{0k}$, which can no longer be expressed in terms of a single Jacobian operator. The second is the presence of the non-colocated Gordon-Haus terms, which will play a more important role.

\section{Spectral Efficiency from the Scattering Data}
\label{sec:SolitonIT}

In the previous section we discussed how we can express the covariance matrix of the Gaussian noise that is injected into the fiber by the amplifiers. In this section we shall use these results to obtain expressions for the mutual information and the corresponding spectral efficiency of the channel. Let us express the output signal as $v(t)\equiv u(t,KL_s)$ and denote the corresponding random process as $V$. Similarly, the input signal is $u(t)\equiv u(t,0)$ with $U$ being the corresponding random process. The mutual information between $U$ and $V$ is
\begin{equation}
\label{MI_def}
I(U;V)=h(V)-h(V|U)
\end{equation}
where $h(V)$ is the entropy of the outgoing signal
\begin{equation}\label{signal_entropyV}
h(V) = -\ex\left[\log_2 p(V)\right]_{U,W}
\end{equation}
with $p(V)$ the probability distribution of the output signal. Similarly, 
\begin{equation}\label{noise_entropyVU_def}
h(V|U) = -\ex\left[\log_2 p(V|U)\right]_{U,W}
\end{equation}
is the entropy of the noise $W$ with distribution $p(V|U)$. Clearly, whatever $p(U)$ is, $p(V|U)$ (and hence $p(V)$) is an extremely complicated functional of the noise, since it includes both non-local insertions of the noise at the amplifiers as well as non-linear propagation following the NLSE. However, due to the integrability of the NLSE, we have seen in the previous sections that the incoming signal $u(t)$ can be mapped into a set of scattering data, which have trivial spatial propagation. In addition, as discussed in Section  \ref{sec:InverseScatteringTransform}, the mapping $U\to X\triangleq[\brho_{in},\bLambda_{in}]$ is a canonical transformation \cite{Faddeev2007_HamiltonianMethods} with Jacobian $J=\frac{\delta U}{\delta X}$  having unit norm determinant, and similarly for the mapping $V\to Y\triangleq [\brho_{out},\bLambda_{out}]$. Consequently, 
\begin{equation}\label{input_entropy}
h(U) = h(X) +\ex\left[\log_2 \det J\right]_{U} =h(X)
\end{equation}
and, correspondingly, $h(V)=h(Y)$ and $h(V|U)=h(Y|X)$. As a result, we have
\begin{equation}\label{MIUV=MIXY}
I(U;V) = I(X;Y)
\end{equation}
To evaluate the noise entropy in the scattering data domain, $h(Y|X)$, we need to consider the fact that the order of solitons is not conserved. This is because the noise affects both the complex eigenvalue $\lambda_n$ and the scattering amplitude $\log b_n$ of each soliton, the real part of the latter corresponding roughly to the center of the soliton in time. Hence all possible assignments from the input solitonic data to the output have to be considered, adding their respective probabilities. Therefore, there is an additional origin of uncertainty in the solitonic degrees of freedom, namely their possible orderings. A similar situation arises in nanoparticle communications \cite{Eckford2007_NanoscaleComms} and the problem of order exchange and communicating with sets in relation to solitons was also noted in \cite{Meron2007_SolitonsCommTheory}. In fact, solitons are particle-like solutions of wave equations, and for pure soliton signals the waveform NLSE channel admits a dual description as a particle communications channel. In contrast, in the continuous part of the spectrum, the (real) eigenvalue is not affected, allowing for the possibility of ordering the continuum modes on the real eigenvalue line. Accounting for orderings, the noise probability in the scattering data domain takes the form
\begin{align}\label{noise_prob_perm}
p(Y|X)= \sum_\pi p_G\left(\pi Y|X\right)
\end{align}
where $p_G(\cdot|\cdot)$ is the Gaussian probability of error described in the previous section with ordered pairs of input and output scattering data, and the sum is over all permutations acting on the solitonic degrees of freedom, such that $\pi Y=[\brho_{out},\pi \bLambda_{out}]$. The corresponding noise entropy is  
\begin{equation}\label{noise_entropyYX_def}
h(Y|X) = -\ex\left[\log_2 \sum_\pi p_G(\pi Y|X)\right]
\end{equation}
where the expectation is over the input distribution $p(X)$ and the (Gaussian) noise distribution $p_G(Y|X)$. Similarly, the output signal entropy can be expressed as
\begin{equation}\label{signal_entropyY}
h(Y) = -\ex\left[\log_2 \ex\left[\sum_\pi p_G(\pi Y|X)\right]_X\right]
\end{equation}

Based on the above, it is easy to show the following inequalities
\begin{align}   \label{MIbounds>}
I(U;V)&\geq h(U) - h_{ord}(Y|X) 
\\
I(U;V)&\leq h_{ord}(Y)-h_{ord}(Y|X) 
\label{MIbounds<}    
\end{align}
where 
\begin{align} \label{hordY}
h_{ord}(Y)=-\ex\left[\log_2 \ex\left[p_G(Y|X)\right]_X\right]_{Y}   
\\
\label{hordYX}
h_{ord}(Y|X)=-\ex\left[\log_2 p_G(Y|X)\right]_{Y,X}
\end{align}
and the latter is the noise entropy without taking into account the ordering uncertainty. The proof of these bounds is given in Appendix \ref{sec:ProofMIBounds}. 

In the absence of the ordering uncertainty, the noise entropy is simply related to the covariance matrix ${\cal S}$ of the additive Gaussian noise, which was the subject of the previous section.  We define $h_G({\cal S})=-\ex\left[\log_2 p_G(Y|X)\right]_{Y,X}$, where $\cal{S}$ is the covariance matrix of the corresponding Gaussian noise. Hence, $h_G({\cal S})$ can be expressed as  
\begin{align}
    \label{GaussianEntropy}
    h_G({\cal S})=\frac{1}{2}\ex\left[\log_2\det({\cal S})\right]_U+M\log_2(\pi e)
\end{align}
where the average is over the input distribution $U$. Here $2M=2BT$ is the number of real degrees of freedom of the signal and is equal to the size of the matrix ${\cal S}$. 

From the above, we can get similar bounds for the performance in bits per channel use. To do this we divide by the number of complex degrees of freedom $M=BT$ and obtain $C=\lim_{T\to \infty}I(U;V)/BT$. Furthermore, in Appendix \ref{sec:ProofMIBounds}, we show that in the low-noise region $\epsilon^2\ll 1$, the two above inequalities in \eqref{MIbounds>}, \eqref{MIbounds<} become tight, so that 
\begin{align}\label{MI_asympt}
    C= \lim_{T\to \infty}\frac{1}{BT} \left[h(U)-h_{G}({\cal S})\right] + o(1)
\end{align}
It should be pointed out that since $h(X)=h(U)$, the entropies of the input in the signal and in the scattering data domain can be used interchangeably. We prefer the above expression, because the constraints on the input signal (such as average power) become thus more transparent.
We thus see that in the low noise limit, the effect of the non-linearities is to color the noise, making the noise-covariance matrix generally non-white. When maximized over the input distribution $p(U)$, subject to an average input power constraint, the above expression gives the spectral efficiency of the system. To leading order in the noise we have
\begin{align}\label{SE_asympt}
\SE = \max_{p(U)} \lim_{T\to \infty}\frac{1}{BT}\left[h(U)-h_G({\cal S})\right] 
\end{align}
Obviously, for any input distribution $p(U)$ we have $\SE\geq C$. Since the entropy of the incoming pulse $h(U)$ is  independent of the fiber-optical channel, in this limit the effect of non-linearities is completely captured by the properties of the covariance matrix ${\cal S}$.

\subsection{Shannon Upper Bound}
\label{sec:Shannon Bound}

An important bound for the fiber spectral efficiency was derived in \cite{yousefi2015upper}, which bounds the spectral efficiency of the non-linear Schroedinger  channel by that of the corresponding linear channel. The bound can be rederived from \eqref{SE_asympt}. 
Indeed: 
\begin{align}\label{Minkowski_ineq}
\det \left( {\cal S}\right)^{\frac{1}{M}}&= 
\det \left( \sum_{k=1}^K{\cal S}_k\right)^{\frac{1}{M}}
\\ \nonumber
 &\geq \left[ \sum_{k=1}^K\det \left({\cal S}_k\right)^{\frac{1}{M}}\right]
\\ \nonumber
 &=K\epsilon^2
\end{align}
The first equality is just the definition of ${\cal S}$, see \eqref{S_mat_def}, while the second line follows from the Minkowski inequality, which relates the determinant of sums of matrices to the sum of their determinants \cite{Hardy_Inequalities_book}. The third line follows from the discussion in Section \ref{sec:PropertiesOfS} and is a manifestation of the fact that the transformation from $\{u(t), u^*(t)\}\to \{\rho_\xi, \lambda, \mu\}$ is a canonical transformation and hence has unit norm determinant. We can now use the fact that the entropy-maximizing signal of $M=BT$ complex degrees of freedom with a variance constraint is an independent Gaussian process, hence
\begin{align}
    \label{GaussianEntropy_inequality}
    h(U)\leq M\log_2 (D\pi e)
\end{align}
with equality when the signal $U$ is Gaussian distributed. Starting from \eqref{SE_asympt}, we thus have
\begin{align}\label{SE_upper_bound}
\SE &\leq \log_2D-\log_2\left[K\epsilon^2\right]
\\ \nonumber
 &= \log_2\left[\frac{P}{K\sigma^2 B}\right]
\end{align}
where we have used the fact that $D=1$ and from this $\epsilon^2=\sigma^2B/P$. This is exactly the small noise limit of the bound derived in \cite{yousefi2015upper}. It states that the optimum spectral efficiency of the additive Gaussian noise model is an upper bound for the corresponding spectral efficiency of the NLSE model with Gaussian noise. We emphasize that a key ingredient of the proof is that the mapping from the signal to the scattering data domain is a canonical transformation, which itself is a consequence of the integrability of the NLSE.

\subsection{A Lower Bound for the Spectral Efficiency}
\label{sec:SELowerBound}

We shall present here a lower bound for the spectral efficiency, which will become useful when discussing the white Gaussian input case. Specifically, we shall start by bounding the expectation of the logarithm of the determinant of the covariance matrix of the noise, as follows:
\begin{align}
\frac{1}{2M} \ex\left[\log_2\det{\cal S}\right]_{U}
&= \ex\left[\log_2\left(\det  {\cal S}\right)^{\frac{1}{2M}}\right]_{U}
\\ \nonumber
&\leq \log_2\ex\left[\left( \det  {\cal S}\right)^{\frac{1}{2M}}\right]_U
\\ \nonumber
&\leq \log_2\left( \det \ex\left[ {\cal S}\right]_U\right)^{\frac{1}{2M}}
\\ \nonumber
&= \frac{1}{2M}\log_2\left( \det \sum_{k=1}^K \ex\left[ {\cal S}_k\right]_U\right)
\end{align}
where the expectation is over the input signal distribution. The first inequality here is due to Jensen, the second due to Minkowski. Therefore, we have  
\begin{align}\label{SE_Gaussian_expression}
\SE &\geq \frac{1}{BT}\left(h(U)-h_G\left(\ex\left[{\cal S}\right] \right)\right) 
\\ \nonumber
&=\frac{1}{2}\log_2 \left(\frac{P}{KB\sigma^2}\right)-\frac{1}{4N}\log_2\det\left(\frac{1}{K}\sum_{k=1}^K\ex\left[{\cal M}_k\right]\right)
\end{align}
Given that the first term is the asymptotic value of the spectral efficiency for linear channels in the low noise limit, we see that the second term is the penalty imposed by the nonlinearity of the channel.

\section{The Gaussian Input Case}
\label{sec:GaussianInputCase}

\begin{figure}[!ht]
\centering{%
\includegraphics[scale=0.5]{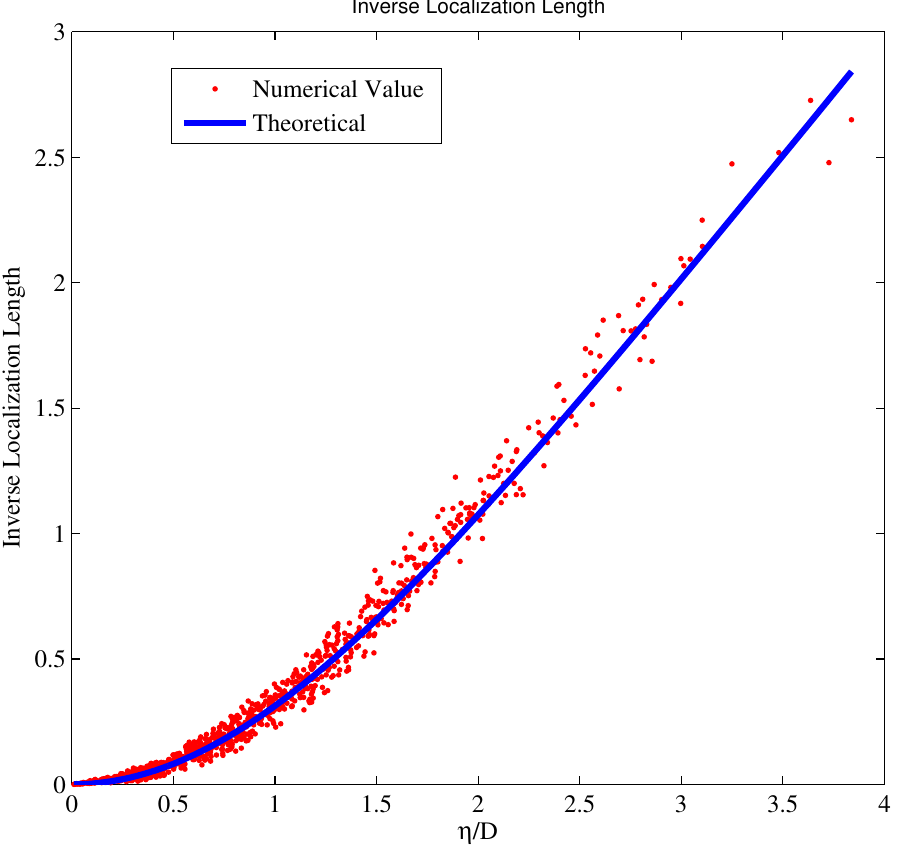}
\caption{Inverse Localization Length as a function of $\eta$. The solid curve is from \eqref{eq:invloclength} in \cite{Kazakopoulos2008_NLSE}. The dots correspond to actual data obtained by calculating the slope of the logarithm of numerically obtained eigenfunctions of the ZS system in \eqref{zs} with delta-correlated input signal $u(t)$.}
\label{fig:InvLocLength}}
\end{figure}

The expression in \eqref{MI_asympt} showcases concretely the mutual information per degree of freedom for the NLSE in the low noise limit. Indeed, one can use it to test the efficiency of different input distributions. One particular case of interest is when the input signal follows a Gaussian distribution, which is the capacity achieving distribution for AWGN. Specifically, we shall assume that the signal is Gaussian with short temporal correlations, namely as in \eqref{u-corrs}. For this type of input, the eigenvalue density of the Zakharov-Shabat operator $\bU$ was calculated analytically in the large bandwidth limit in \cite{Kazakopoulos2008_NLSE}, exhibiting zero density of real eigenvalues. 
As mentioned earlier, the complex eigenvalues may be assumed to be mostly independent from each other due to orthogonality of their corresponding eigenfunctions. In addition, since these are localized (solitonic), their overlap is limited and hence we expect that the covariance matrix ${\cal S}$, consisting of overlap integrals between these eigenfunctions, \eqref{S_mat_def}, will generally be sparse. Of course, this overlap depends on the inverse localization length, whose dependence on $\lambda$ was calculated analytically in \cite{Kazakopoulos2008_NLSE} and takes the form:
\begin{eqnarray}
\kappa(\lambda) = \frac{D}{2}\left[\frac{2\eta}{D}\coth\left( \frac{2\eta}{D} \right) -1\right]
\label{eq:invloclength}
\end{eqnarray}
where $\eta$ is the imaginary part of $\lambda$, and $D=1$. In Fig. \ref{fig:InvLocLength} we plot the above analytic result, and show that it is in good agreement with numerically generated solutions of the ZS system. The figure also showcases that no finite density of real eigenvalue (extended) states exists, at least in this high bandwidth limit. 

We shall now provide an independent approach to relate the number of d.o.f. in the input signal, $M=BT$, with the number of solitons for specific case of Gaussian inputs. For a given signal $u(t)$, the total energy $E(\{u(t)\})$ is related to the imaginary parts of the eigenvalues as follows \cite{Konotop_book}:
\begin{align}\label{energy}
 E(\{u(t)\})\triangleq\int_0^{T_s} |u(t)|^2dt \geq 4\sum_{k=1}^N \eta_k 
\end{align}
where $N$ is the number of soliton quadruplets (see  \cite[{Section 3.2.2}]{Ablowitz2004_DNLS_book} with only solitons with positive real and imaginary parts entering the sum and the inequality becoming an equality in the absence of real eigenvalues. 
For large times, the integral is approximately equal to its average, which from \eqref{u-corrs} can be found to be equal to $DBt_sT_s=DBT=BT$. The right-hand side of the above equation becomes asymptotically equal to $4N\ex[\eta]$, which from \cite{Kazakopoulos2008_NLSE} can be evaluated to be $2ND$. As a result, we have $N\leq BT/2$. Since the density of real eigenvalues has been shown in \cite{Kazakopoulos2008_NLSE} to be zero, and additionally, as seen in Fig. \ref{fig:InvLocLength}, we have found no real eigenvalues numerically, we shall assume in this case that $N=BT/2$.

For white Gaussian input with statistics given by \eqref{u-corrs}, we can benefit from a key simplification. In particular, given that the total energy (see left hand side in \eqref{energy}) is conserved during the propagation down the fiber (in the low noise limit) and that the distribution of the Gaussian input signal is proportional to $\exp\left(-a E(\{u(t)\})\right)$, where $a$ is a constant, we conclude that the distribution itself is invariant during propagation. Hence, the expectation of the covariance matrix at the $k$th fiber segment, $\ex\left[{\cal S}_k\right]_U$ depends on $k$ only through $\alpha_k$. Therefore, we can take advantage of the lower bound discussed in Section \ref{sec:SELowerBound} and evaluate the expectation $\ex[{\cal M}^k]$ over initial Gaussian pulses, without having to resort to propagation of the scattering data down the fiber.

\section{Numerical Evaluation of Spectral Efficiency}
\label{sec:Numerics}

To obtain a quantitative estimate of the mutual information per degree of freedom for the Gaussian input channel we shall use \eqref{MI_asympt}. The entropy of the input signal can be readily evaluated using \eqref{GaussianEntropy_inequality} with an equality sign. To calculate the noise entropy $h_G({\cal S})$ we need to to evaluate the covariance matrices ${\cal M}^k$  of the noise injected at amplifiers $k=1, 2, \ldots,K$ appearing in \eqref{S_mat_def}. To do this we have to propagate the random input signal $u(t)$ down the fiber up to distance $x=KL_s$. For concreteness, we have assumed periodic boundary conditions on $u(t)$ \cite{kamalian2016periodicII}, as is customary when one wants to minimize the effects of boundaries due to finite size pulses. As discussed in Section \ref{sec:ChannelModel} we have chosen the parameters $R,t_s$, such that $D=1$, thus with input signal of unit norm and, for $Bt_s\gg 1$, uncorrelated. However, this scaling will also make the distance between amplifiers $L_s$ power dependent.

\begin{table}[ht]
\begin{tabular}{|c|c|c|}
\hline
\textbf{Parameter}  & \textbf{Symbol} & \textbf{Value} 
\\ \hline 
GVD &$\beta_2$  & 21.67ps$^2$/km  
\\ \hline 
Nonlinearity & $\gamma$  & 1.27W$^{-1}$km$^{-1}$ 
\\ \hline 
Bandwidth & $B$  & 100GHz   
\\ \hline 
Amplifier Distance & $L$  & 100km 
\\ \hline 
Maximum Distance & $K_{max}L$  & 2000km   
\\ \hline 
Photon Energy & $E_{ph}$   & $13.2\cdot10^{-20}$J
\\ \hline 
Emission Factor&  $n_{sp}$  & 1
\\ \hline 
Attenuation & $\alpha_{loss}$  & 0.2dB/km
\\ \hline 
Noise & $\sigma^2$  & $n_{sp}E_{ph}(e^{\alpha_{loss}L}-1)$ 
\\ \hline 
\end{tabular}
\caption{Parameter values (\cite{essiambre2010capacity}) for the simulations in Section \ref{sec:Numerics}.}
\label{tab:parameters}
\end{table}

A challenge we had to address is that, due to the ``rough'', uncorrelated nature of the incoming signal, traditional methods of  numerically solving the non-linear Schroedinger equation, such as the split-Fourier method and its variants, fail to keep the values of the complex eigenvalues of the Zakharov-Shabat system constant, reminiscent of soliton ''turbulence'', discussed in  \cite{Suret2024_SolitonGas}. Instead, we used discretized symplectic methods \cite{Tang2007_SymplecticMethodsALDiscreteNLSE}, which take advantage of the fact that the discrete Ablowitz-Ladik equation is integrable and hence its evolution can be seen as a canonical transformation. While significantly more time-consuming, this method ensured near-constancy of the eigenvalue locations, an indication that indeed the numerical evolution remained accurate. At every distance where an amplifier is located, we used the propagated signal $u(t,kL_s)$, for $k=1,\ldots,K$ as input into the Zakharov-Shabat operator to calculate the eigenfunctions $\bPsi_{n}(t,kL_s)$, for $n=1,\ldots,N$ employing the modified Ablowitz-Ladik scheme \cite{Weideman_Herbst_Modified_Ablowitz_Ladik}). In addition, the  derivatives of the eigenfunctions $\bPsi_{n}'(t,kL_s)$, appearing in \eqref{dlogb_du}, were evaluated from $\bPsi_n$ and $\lambda_n$ using the equation:
\begin{equation}\label{psi_derivative}
\left(\bU-\lambda_n\right)\bPsi_n'=\bPsi_n
\end{equation}
Having $\bPsi_n$ and $\bPsi_n'$, we obtained the covariance matrices ${\cal M}^k$ as discussed in Appendix \ref{sec:PertTheory}, and then, by adding them up the noise covariance matrix ${\cal S}$ can be calculated.


\begin{figure}[ht]
\centering{%
\includegraphics[scale=0.35]{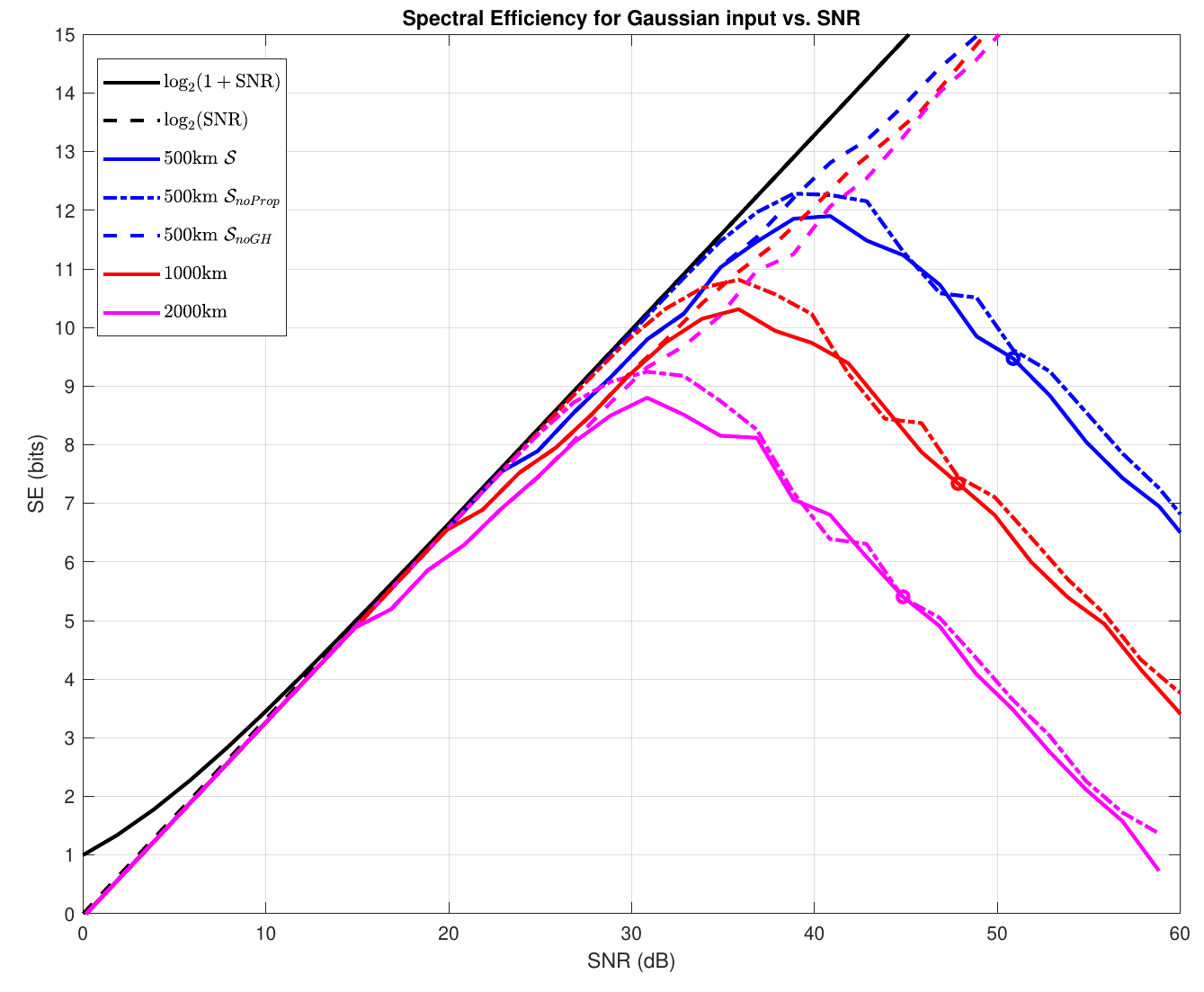}
\caption{Spectral Efficiency curves as a function of SNR.  The black curves are the linear Shannon limit, depicted here for comparison (see \eqref{SE_upper_bound}). The three sets of colored curves correspond to three total distances, namely $500$km (blue), $1000$km (red), $2000$km (magenta). For each distance, the solid curve corresponds to the spectral efficiency  
with the full covariance matrix ${\cal S}$ in \eqref{S_mat_def}. The dashed curve uses the covariance matrix ${\cal S}_{noGH}$ in \eqref{eq:SnoGH}, which does not include the effect of Gordon-Haus terms, and hence continuously increases with power. The dotted curve corresponds to the simplified covariance matrix $\SnoProp$ in \eqref{eq:SsimpleGH}, which still follows the solid curves very closely.
Parameter values from Table \ref{tab:parameters} have been used in the simulations.
}
\label{fig:SE_vs_SINR}
}
\end{figure}

\subsection{Analysis of Results}

In Fig. \ref{fig:SE_vs_SINR} we present the spectral efficiency for white complex Gaussian inputs, evaluated using $BT=2N=200$, and taking into account only a single polarization of the light signal. Table \ref{tab:parameters} summarizes the commonly accepted \cite{essiambre2010capacity} parameter values we have also used. Since all eigenvalues of the input are solitonic (complex), we did not have to analyze the continuous degrees of freedom of the system. For comparison, the figure also includes the (black) curves for $\log_2(1+\SNR)$ and $\log_2(\SNR)$,  as benchmarks, confirming that the Shannon limit discussed in Section \ref{sec:Shannon Bound} and in \cite{yousefi2015upper} is indeed an upper bound for the spectral efficiency. We plot the performance of the system for three distances $L_{tot}=$ 500, 1000 and 2000km, with identical amplifiers every 100km. All three (solid) curves have the same qualitative behavior: They initially rise, closely following the Shannon limit, reaching a peak at a certain SNR, beyond which they fall. This peak spectral efficiency is distance-dependent, as expected, being higher for shorter distances, and is somewhat higher than the numerical and semi-analytical values obtained in the literature using other methods \cite{essiambre2012capacity, mecozzi2012nonlinear}.
The small circles appearing on the curves indicate the SNR value (correspondingly the input power) beyond which the condition $Bt_s\gg 1$ no longer holds. Hence the behavior to the right of these circles is not to be trusted. However, this is not a problem, since these points are to right of the peak of the curves and hence not of particular interest. It should also be pointed out that due to the complexity of the calculation, all curves correspond to a single realization of the initial pulse $u(t,x=0)$ rather than an average as \eqref{GaussianEntropy} indicates. Nevertheless, initial calculations with shorter pulses have shown that the value of $\log_2(\det({\cal S}))$ has negligible fluctuations, essentially self-averaging for large sizes. 

As mentioned in Section \ref{sec:PropertiesOfS} there are two aspects of the covariance matrix ${\cal S}$ which depart from the i.i.d. linear additive Gaussian noise covariance matrix. The first has to do with the sum of matrices ${\cal M}^{0k}$ for each amplifier $k=1,\ldots,K$. While separately each ${\cal M}^{0k}$ has unit determinant, the determinant of their sum can in principle take on any value. To test the effect of these matrices on the outcome, we plot the spectral efficiency including only these terms in the covariance matrix, and omitting the terms that give rise to the Gordon-Haus effect, i.e. using 
\begin{equation}\label{eq:SnoGH}
\overline{{\cal S}}_{noGH}=\epsilon^2 \sum_{k=1}^K {\cal M}^{0k}
\end{equation}
Surprisingly, we get the (dotted) curves, which, despite a small reduction in value, follow the linear Shannon limit, increasing linearly with $\log(\SNR)$. This indicates that it is the Gordon-Haus effect that is responsible for the eventual decline of the spectral efficiency curves as a function of SNR.

The Gordon-Haus effect acts essentially by randomly shifting the velocities of the solitonic degrees of freedom. To understand how much the spreading is affected by the modification of the pulse through the evolution of the eigenfunctions, as it propagates down the fiber, we performed a very simple test, namely we generated the eigenfunctions directly from a white Gaussian input pulse and used them to obtain the covariance matrices ${\cal M}^k$ without any propagation. In this case these matrices depend on $k$ only through the parameter $\alpha_k$ in \eqref{eq:Dk_def}. After some algebra, we obtain  
\begin{align}\label{eq:SsimpleGH}
\SnoProp &=K\epsilon^2 
\left[
  \begin{array}{ll}
    {\cal M}_{\lambda \lambda}^{0}  &  {\cal M}_{\lambda\mu}^{0\dagger} \\
    {\cal M}_{\lambda\mu }^{0} &   {\cal M}_{\mu \mu}^{0} + 
    \zeta_K {\cal D}_1{\cal M}^0_{\lambda\lambda}{\cal D}_1^\dagger
  \end{array}
\right]
\end{align}
where
\begin{equation}
\zeta_K =\sum_{k=1}^K\frac{\alpha_k^2}{K}-\left(\sum_{k=1}^K\frac{\alpha_k}{K}\right)^2=\frac{16}{3}(K^2-1)L_s^2
\end{equation}
The spectral efficiency curves (dashed) obtained this way and depicted in Fig. \ref{fig:SE_vs_SINR}, show a behavior similar to the curves obtained with the proper covariance matrix ${\cal S}$. We thus see that the variations in the eigenfunctions due to propagation down the fiber play a minor role in the appearance of the plateau in spectral efficiency. Rather, the Gordon-Haus effect appearing through the parameters $\alpha_k$ is the culprit. 
It is worth noting that the simplified form of $\SnoProp$ showcases the way in which the Gordon-Haus term (second in the lower-right block) enters the covariance matrix. More specifically, it shows that for small values of $KL_s=\frac{KL(\gamma P)^2}{\beta_2 B^2}$, the spectral efficiency is approximately equal to $\SE\approx \log_2(\SNR)$, following the linear Shannon curve, while for large values of the parameter $KL_s$, the Gordon-Haus effect dominates and the spectral efficiency curve behaves as 
\begin{align}
    \SE&\approx \log_2\left(\frac{P}{KB\sigma^2}\right)-\log_2\left(\frac{KL(\gamma P)^2}{\beta_2 B^2}\right) 
    \\ \nonumber
    &\approx\log_2\left(\frac{\beta_2 B^2}{(\gamma K\sigma^2 B)KL(\gamma P)}\right)
\end{align}
Since the behavior of the curves appearing in Fig. \ref{fig:SE_vs_SINR} with $\SnoProp$ in \eqref{eq:SsimpleGH} is very similar to the full covariance matrix curves, we expect the above interpretation of the behavior of the spectral efficiency to hold for those as well.


\section{Discussion and Outlook}
\label{sec:Conclusion}

Underlying the challenge to understand the fundamental limits of communications through non-linear optical fibers is the competition between dispersion and non-linearity, which complicates the effects of noise injected during amplification. Taking the low-noise limit provides the opportunity to study more closely the interplay between these effects. In the above analysis we have taken advantage the integrability of the NLSE in this limit to obtain an expression for the spectral efficiency of the system, expressed using the entropy of the input signal and the covariance matrix of the distributed Gaussian noise. We showed that the properties of the latter, which can be expressed solely in terms the scattering data obtained from the non-linear Fourier transform of the input signal, are tied to the integrability of the NLSE. 

This throughput expression, valid in principle for any input distribution, allows for the evaluation of the corresponding spectral efficiency and the maximization over the incoming signal distribution. Taking advantage of these properties we have re-established within this framework the Shannon upper bound of the spectral efficiency first derived in \cite{yousefi2015upper}. More importantly, emerging from the structure of the covariance matrix is the relevance of the Gordon-Haus effect and its role in dampening the increase of the throughput with power. In fact, we saw both numerically, but also from the qualitative analysis of the covariance matrix, that while initially the spectral efficiency increases in step with the linear Shannon case, beyond a certain maximum value, the spectral efficiency starts decreasing. While this behavior has been seen in most capacity analyses of the NLSE \cite{Mitra_Stark_2001,Essiambre2010_CapacityLimitsOpticalFiberNetworks, derevyanko2016capacity}, this is the first time it is unambiguously shown that the Gordon-Haus effect is responsible for this phenomenon, which for increasing power destroys the ordering of the soliton modes of the signal. 

It can be argued that the above results are tied to the Gaussian distribution we have assumed for the input signal. However, it should be noted that in the high-bandwidth (``white-noise'') input-power limit that we have used (see \eqref{u-corrs}), any input distribution approaches the white-Gaussian statistics, with the remaining degree-of-freedom being the distribution of the total average power $P$, which of course is optimal if chosen at the peak of the corresponding spectral efficiency curve.

Further, it is true that the input distribution we have used in the numerics does not have a meaningful density of non-solitonic (continuous) modes, which do not exhibit the Gordon-Haus effect, but also have a maximum in the spectral efficiency curve \cite{derevyanko2016capacity}. It would therefore be interesting to generalize our analysis for input distributions with both solitonic and non-solitonic degrees of freedom.

In addition, as also mentioned in the Introduction, it should be pointed out that the above analysis and results are not in contrast with the results for dispersionless channels, which correspond to $\beta_2=0$ and hence $Bt_s=0$. However, this limit of dispersionless channels is singular, because, while $\beta_2=0$, the bandwidth is essentially infinite, with independent signals at every symbol. Hence the product $\beta_2 B^2$ corresponding to the inverse length scale of the dispersion term is ill-defined. Therefore, a more careful perturbative  analysis for small $\beta_2$ should be conducted to ascertain if this result can hold within the fully integrable NLSE scheme. 

Finally, while the above discussion points to the hypothesis that the peaked spectral efficiency behavior is universal, it is still possible, in the opposite limit we analyzed, namely $Bt_s\ll 1$, that the spectral efficiency will increase with no bound with power. To this end we believe that our derived spectral efficiency expression can be useful as well in other integrable systems that share the same qualitative characteristics, such as the defocusing NLSE \cite{yousefi2019linear}, the Manakov equation (two-polarization NLSE) \cite{goossens2017polarization} or the newly proposed multimode fibers \cite{essiambre2013breakthroughs}.

\appendix

\section{Noise Covariance Matrix}
\label{sec:PertTheory}

We now evaluate the covariance matrix of the Gaussian noise. Focusing on a single amplifier $k$, we define the covariance matrices
\begin{align}\label{blambdamublambdamu1}
 {\cal M}^{0k}_{\lambda\lambda} &= 
\left[
  \begin{array}{cc}
    {\cal M}^{0k}_{\lambda 1} & {\cal M}^{0k}_{\lambda 2} \\
    \left({\cal M}^{0k}_{\lambda 2}\right)^* & \left({\cal M}^{0k}_{\lambda 1}\right)^*
  \end{array}
\right] 
\\ \notag
 {\cal M}^{0k}_{\rho\rho} &=
 \left[
  \begin{array}{cc}
    {\cal M}^{0k}_{\rho 1} & {\cal M}^{0k}_{\rho 2} \\
    \left({\cal M}^{0k}_{\rho 2}\right)^* & \left({\cal M}^{0k}_{\rho 1}\right)^*
  \end{array}
\right] 
\\ \notag
 {\cal M}^{0k}_{\mu\mu} &=
\left[
  \begin{array}{cc}
    {\cal M}^{0k}_{\mu 1} & {\cal M}^{0k}_{\mu 2} \\
    \left({\cal M}^{0k}_{\mu 2}\right)^* & \left({\cal M}^{0k}_{\mu 1}\right)^*
  \end{array}
\right] 
\end{align}
with their corresponding block matrices having elements 
\begin{align}\label{X_covariance1_diag}
&{\cal M}^{0k}_{\lambda 1, nm} = \ex\left[\delta\lambda_{n}^{0} \delta\lambda_{m}^{0*} \right]
\\ \notag
&= \int \frac{\psi_{2n,k}^2(t) \psi_{2m,k}^{*2}(t)+\psi_{1n,k}^2(t) \psi_{1m,k}^{*2}(t)}{\gamma_{n,k}^{}\gamma_{m,k}^{*}} dt
\\ \notag
&{\cal M}^{0k}_{\lambda 2, nm} = \ex\left[ \delta\lambda_{n}^{0} \delta\lambda_{m}^{0} \right] 
\\ \notag
&=-\int \frac{\psi_{1n,k}^2(t) \psi_{2m,k}^{2}(t)+\psi_{2n,k}^2(t) \psi_{1m,k}^{2}(t)}{\gamma_{n,k}^{}\gamma_{m,k}^{}} dt
\\ \notag
&{\cal M}^{0k}_{\rho 1, \xi \xi'} = \ex\left[\delta\rho_\xi^0\delta\rho_{\xi'}^{0*}\right]
\\ \notag
&=\int \frac{\phi_{2\xi,k}^2(t) \phi_{2\xi',k}^{*2}(t)+\phi_{1\xi,k}^2(t) \phi_{1\xi',k}^{*2}(t)}{a^2(\xi)a^{*2}(\xi')} dt
\end{align}

\begin{align*}
\\ \notag
&{\cal M}^{0k}_{\rho 2, \xi\xi'} = \ex\left[\delta\rho^0_\xi\delta\rho_{\xi'}^0\right]
\\ \notag
&=\int \frac{\phi_{2\xi,k}(t)^2 \phi_{1\xi',k}(t)^2+\phi_{1\xi,k}(t)^2 \phi_{2\xi',k}^{*}(t)^2}{a^2(\xi)a^{2}(\xi')} dt
\\ \notag
&{\cal M}^{0k}_{\mu 1, nm} = \ex\left[\delta\mu_n^{}\delta\mu_m^{*}\right]
\\ \notag
&= \int \frac{\left(\psi_{2n,k}^2(t)\right)' \left(\psi_{2m,k}^{*2}(t)\right)'+\left(\psi_{1n,k}^2(t)\right)' \left(\psi_{1m,k}^{*2}(t)\right)'}{\gamma_{n,k}^{}\gamma_{m,k}^{*}} dt
\\ \notag
&{\cal M}^{0k}_{\mu 2, nm} = \ex\left[\delta\mu_n^{}\delta\mu_m^{}\right]
\\ \notag
&= \int \frac{\left(\psi_{2n,k}^2(t)\right)' \left(\psi_{1m,k}^{2}(t)\right)'+\left(\psi_{1n,k}^2(t)\right)' \left(\psi_{2m,k}^{2}(t)\right)'}{-\gamma_{n,k}^{}\gamma_{m,k}^{}} dt
\end{align*}
and
\begin{align}\label{blambdamublambdamu2}
 {\cal M}^{0k}_{\lambda\rho} &=
\left[
  \begin{array}{cc}
    {\cal M}^{0k}_{\lambda\rho 1} & {\cal M}^{0k}_{\lambda\rho 2} \\
    \left({\cal M}^{0k}_{\lambda\rho 2}\right)^* & \left({\cal M}^{0k}_{\lambda\rho 1}\right)^*
  \end{array}
\right] 
\\ \notag
 {\cal M}^{0k}_{\lambda\mu} &=
\left[
  \begin{array}{cc}
    {\cal M}^{0k}_{\lambda\mu 1} & {\cal M}^{0k}_{\lambda\mu 2} \\
    \left({\cal M}^{0k}_{\lambda\mu 2}\right)^* & \left({\cal M}^{0k}_{\lambda\mu 1}\right)^*
  \end{array}
\right] 
\\ \notag
 {\cal M}^{0k}_{\rho\mu} &=
\left[
  \begin{array}{cc}
    {\cal M}^{0k}_{\rho\mu 1} & {\cal M}^{0k}_{\rho\mu 2} \\
    \left({\cal M}^{0k}_{\rho\mu 2}\right)^* & \left({\cal M}^{0k}_{\rho\mu 1}\right)^*
  \end{array}
\right] 
\end{align}
with their corresponding block matrices composed of the elements 
\begin{align}\label{X_covariance2_offdiag}
&{\cal M}^{0k}_{\lambda\rho 1, n\xi} = \ex\left[\delta\lambda_{n}^{0}\delta\rho_\xi^{0*}\right]
\\ \notag
&= \int \frac{\psi_{2n,k}^2(t) \phi_{2\xi,k}^{*2}(t)+\psi_{1n,k}^2(t) \phi_{1\xi,k}^{*2}(t)}{-i\gamma_{n,k}^{}a_{k}^{*}(\lambda)^2} dt
\\ \notag
&{\cal M}^{0k}_{\lambda\rho 2, n\xi} = \ex\left[\delta\lambda_{n}^{0}\delta\rho^{0}_\xi\right]
\\ \notag
&= \int \frac{\psi_{2n,k}^2(t)\phi_{1\xi,k}^{2}(t)+\psi_{1n,k}^2(t) \phi_{2\xi,k}^{2}(t)}{-i\gamma_{n,k}^{}a_{\xi,k}^2} dt
\\ \notag
&{\cal M}^{0k}_{\rho\mu 1, \xi n} = \ex\left[\delta\rho^{0}_\xi\delta\mu_n^{*}\right]
\\ \notag
&= \int \frac{\phi_{1\xi,k}^{2}(t) \left(\psi_{1n,k}^{*2}(t)\right)' + \phi_{2\xi,k}^{2}(t) \left(\psi_{2n,k}^{*2}(t)\right)' }{ia_{\xi,k}^2\gamma_{n,k}^{*}} dt
\\ \notag
&{\cal M}^{0k}_{\rho\mu 2, \xi n} =  \ex\left[\delta\rho_\xi^{0}\delta\mu_n^{}\right]
\\ \notag
&= \int \frac{\phi_{1\xi,k}^{2}(t) \left(\psi_{2n,k}^{2}(t)\right)' + \phi_{2\xi,k}^{2}(t) \left(\psi_{1n,k}^{2}(t)\right)' }{-ia_{k}(\lambda)^2\gamma_{n,k}^{}} dt
\end{align}
\vspace{1cm}
\begin{align*}
\\ \notag
&{\cal M}^{0k}_{\lambda\mu 1, nm} = \ex\left[\delta\lambda_{n}^{0}\delta\mu_m^{*}\right]
\\ \notag
&= \int \frac{\psi_{2n,k}^2(t) \left(\psi_{2m,k}^{*2}(t)\right)'+\psi_{1n,k}^2(t) \left(\psi_{1m,k}^{*2}(t)\right)'}{\gamma_{n,k}^{}\gamma_{m,k}^{*}} dt
\\ \notag
\end{align*}
\begin{align*}
&{\cal M}^{0k}_{\lambda\mu 2, nm} =  \ex\left[\delta\lambda_{n}^{0}\delta\mu_m^{}\right]
\\ \notag
&= \int \frac{\psi_{2n,k}^2(t)\left(\psi_{1m,k}^{2}(t)\right)'+\psi_{1n,k}^2(t) \left(\psi_{2m,k}^{2}(t)\right)'}{-\gamma_{n,k}^{}\gamma_{m,k}^{}} dt
\end{align*}



\section{Proof of Eqs. \eqref{MIbounds>}, \eqref{MIbounds<} and \eqref{MI_asympt} }
\label{sec:ProofMIBounds}

The lower bound in \eqref{MIbounds>} is a consequence of the inequalities $h(V)\geq h(U)$ and the fact that 
\begin{align}
&h(Y|X)= -\ex_{X,Y} \log_2\left[\sum_\pi p_G(\pi Y|X)\right] 
 \\ \nonumber
 &\leq -\ex_{X,Y} \log_2\left[ p_G(Y|X)\right]
=h_{ord}(Y|X)   
\end{align}

To prove the upper bound in \eqref{MIbounds<}, we start by separating the integration over $Y$ to an integration over the subspace, 
in which the solitonic degrees of freedom have a specific ordering, denoted by $\vec{Y}$ and a sum over permutations of this ordering, in such a way that any $Y$ can be written in a unique way as $Y=\pi\vec{Y}$ and $\ex_{Y}[f]=\ex_{\pi,\vec{Y}}[f]=\int_{\vec{Y}}\sum_\pi p(\pi \vec{Y}|X) f$. 
We then have from Jensen's inequality
\begin{align}
&I(X,Y)-(h_{ord}(Y)-h_{ord}(Y|X)) = 
\\ \nonumber
&\ex_{X,Y}\left[\log_2\left(\frac{\ex_{\pi'}\left[p_G(\pi'Y|X)\right]\ex_{X'}\left[p_G(Y|X')\right]}{\ex_{X',\pi'}\left[p_G(\pi' Y|X')\right]p_G(Y|X)}  \right)\right]
    \\ \nonumber
    &\leq \log_2\left(\ex_{X,Y}\left[\frac{\ex_{\pi'}\left[p_G(\pi'Y|X)\right]\ex_{X'}\left[p_G(Y|X')\right]}{\ex_{X',\pi'}\left[p_G(\pi' Y|X')\right]p_G(Y|X)} \right]\right)
        \\ \nonumber
    &= \log_2\left(\ex_{X}\int_{\vec{Y}} \left[\frac{\ex_{\pi'}\left[p_G(\pi'\vec{Y}|X)\right]\ex_{X',\pi}\left[p_G(\pi \vec{Y}|X')\right]}{\ex_{X',\pi'}\left[p_G(\pi' \vec{Y}|X')\right]} \right]\right)
    \\ \nonumber
    &= \log_2\left(\ex_{X}\int_{\vec{Y}} \ex_{\pi'}\left[p_G(\pi'\vec{Y}|X)\right]\right) = 0
\end{align}
from which \eqref{MIbounds<} follows. 

To prove \eqref{MI_asympt}, we need to show that $h_{ord}(Y)\approx h(X)$ in the low noise limit. In this limit, $p_G(Y|X)\approx \delta(Y-X)$ in the sense that the scattering data in the presence of vanishing noise stay asymptotically close to their original values. Hence, $\ex_X[p_G(Y|X)]\approx p_X(Y)$, where $p_X(\cdot)$ is the distribution of the incoming scattering data. Therefore, $h_{ord}(Y)\approx h(X)=h(U)$.  

\bibliographystyle{apsrev4-2}
%


\end{document}

%% file: macros_alm.tex
\usepackage{amsmath}
\usepackage{amssymb}
\usepackage{amsfonts}
\usepackage{amsthm}

\usepackage{amsbsy}
\usepackage{times}
\usepackage{color}
\usepackage{enumerate}

\usepackage{graphicx}
\usepackage{epsfig}
\usepackage{bm}
\usepackage[normalem]{ulem}

\usepackage{soul}

\usepackage[svgnames]{xcolor}
\usepackage{subfigure}

\usepackage{acronym}
\usepackage{latexsym}

\usepackage[sort&compress]{natbib}

\newcommand{\N}{\mathbb{N}}

\DeclareMathOperator{\ex}{\mathbb{E}}

\theoremstyle{plain}

\newtheorem*{corollary*}{Corollary}

\theoremstyle{definition}

\newtheorem*{definition*}{Definition}

\theoremstyle{remark}

\newtheorem*{remark*}{Remark}

\newcommand{\bA}{\mathbf{A}}
\newcommand{\bB}{\mathbf{B}}
\newcommand{\bC}{\mathbf{C}}
\newcommand{\bD}{\mathbf{D}}

\newcommand{\bU}{\mathbf{U}}

\newcommand{\bDelta}{\mathbf{\Delta}}

\newcommand{\bLambda}{\mathbf{\Lambda}}

\newcommand{\bPsi}{\mathbf{\Psi}}

\newcommand{\blambda}{\boldsymbol{\lambda}}
\newcommand{\bmu}{\boldsymbol{\mu}}

\newcommand{\brho}{\boldsymbol{\rho}}

\def\vec#1{\overrightarrow{\mathbf{#1}} } 

\newfont{\bb}{msbm10 scaled 1100}